\documentclass{birkjour}

\usepackage[T2A]{fontenc}
\usepackage[english]{babel}
\usepackage{latexsym}
\usepackage{amssymb}
\usepackage{amsmath}

 \newtheorem{thm}{Theorem}[section]
\def\proof{\medbreak\noindent{\bf Proof}}
\def\theorem #1. #2\par{\medbreak
  \noindent{\tt {\bf Theorem #1.}\enspace}{\sl#2\par}%
  \ifdim\lastskip<\medskipamount \removelastskip\penalty55\medskip\fi}
\def\lemma #1. #2\par{\medbreak
  \noindent{\tt {\bf Lemma #1.}\enspace}{\sl#2\par}%
  \ifdim\lastskip<\medskipamount \removelastskip\penalty55\medskip\fi}

\def\{{\lbrace}
\def\}{\rbrace}

\def\cl{{C}\!\ell}

\def\tr{{\rm tr}}
\def\R{{\Bbb R}}
\def\C{{\Bbb C}}

\def\L{{\cal L}}

\def\diag{{\rm diag}}
\def\Even{{\rm Even}}
\def\Odd{{\rm Odd}}

\def\O{{\rm O}}
\def\o{{\rm so}}
\def\Sp{{\rm Sp}}
\def\sp{{\rm sp}}
\def\Spin{{\rm Spin}}

\def\be{\begin{equation}}
\def\ee{\end{equation}}

\def\Mat{{\rm Mat}}

\newcommand{\GL}{{\rm GL}}
\newcommand{\gl}{{\rm gl}}

\begin{document}

\title[Symplectic, orthogonal and linear Lie groups in Clifford algebra]{Symplectic, orthogonal and linear Lie groups in Clifford algebra}

\author[D.~S.~Shirokov]{D.~S.~Shirokov}
\address{
1. A. A. Kharkevich Institute for Information Transmission Problems, Russian Academy of Sciences,\\
Bolshoy Karetny per. 19, 127994, Moscow, Russia;\\
2. N. E. Bauman Moscow State Technical University,\\
ul. Baumanskay 2-ya, 5, 105005, Moscow, Russia.}
\email{dm.shirokov@gmail.com}
\begin{abstract}
In this paper we prove isomorphisms between 5 Lie groups (of arbitrary dimension and fixed signatures) in Clifford algebra and classical matrix Lie groups - symplectic, orthogonal and linear groups. Also we obtain isomorphisms of corresponding Lie algebras.
\end{abstract}
\subjclass{15A66}
\keywords{Clifford algebra, symplectic group, orthogonal group, linear group, Lie group, Lie algebra, spin group}
\maketitle

\section{Introduction}

In this paper we further develop results of the paper \cite{MarchukDyab}. We prove isomorphisms between 5 Lie groups in Clifford algebra formalism (see (\ref{grou}))
$$\Sp\O^\R_{2i1}\cl(p,q),\,
\Sp\O^\R_{2i3}\cl(p,q),\,
\Sp\O^\R_{23}\cl(p,q),\,
\Sp\O^\R_{12}\cl(p,q),\,
\Sp\O^\R_{2}\cl(p,q)$$
and classical matrix groups (Theorem 4.2). Also we obtain statements for corresponding Lie algebras.

Note that in \cite{MarchukDyab} you can find the proof of special case of the statement from this paper about isomorphism $$\Sp\O^\R_{2i1}\cl(1,3)=\{U\in\cl^\R_\Even(1,3)\oplus i\cl^\R_\Odd(1,3) : U^\ddagger U=e\}\simeq$$
$$\simeq\Sp(2,\R)=\{A\in\Mat(4, \R): A^T\Omega A=\Omega\}.$$
In this paper we prove isomorphisms for arbitrary dimension. We consider 5 different groups in the cases of several signatures.

Note that group $\Sp\O^\R_{2i1}\cl(p,q)$ is a subgroup of pseudo-unitary group $W_{23i01}\cl(p,q)=\{U\in \cl(p,q)\, | \, U^\ddagger U=e\}$ (see \cite{MarchukDyab}, \cite{pseudou}, \cite{Snygg}).

Also note, that spin group $\Spin_{+}(p,q)$ is a subgroup of all 5 considered Lie groups. Moreover, group $\Spin_{+}(p,q)$ coincides with group $\Sp\O^\R_{2}\cl(p,q)$ in the case of dimensions $n\leq 5$.

\section{Clifford algebras and quaternion types}

Consider complex Clifford algebra $\cl(p,q)$ (or real $\cl^\R(p,q)$) with $p+q=n$, $n\geq1$.
The construction of Clifford algebra is discussed in details in \cite{Lounesto} or \cite{Marchuk:Shirokov}.

Let $e$ be the identity element and let $e^a$, $a=1,\ldots,n$ be generators of the Clifford algebra $\cl(p,q)$,
$$
e^a e^b+e^b e^a=2\eta^{ab}e,
$$
where $\eta=||\eta^{ab}||$ is the diagonal matrix with $p$ pieces of $+1$ and $q$ pieces of $-1$ on the diagonal. Elements
$$
e^{a_1\ldots a_k}=e^{a_1}\ldots e^{a_k},\quad a_1<\ldots<a_k,\,k=1,\ldots,n,
$$
together with the identity element $e$, form the basis of the Clifford
algebra. The number of basis elements is equal to $2^n$.

Any Clifford algebra element $U\in\cl(p,q)$ can be written in the form
\begin{eqnarray}
U=ue+u_a e^a+\sum_{a_1<a_2}u_{a_1 a_2}e^{a_1 a_2}+\ldots+u_{1\ldots n}e^{1\ldots n},\label{U}
\end{eqnarray}
where $u, u_a, u_{a_1 a_2}, \ldots, u_{1\ldots n}$ are complex (real) numbers.

We denote by $\cl_k(p,q)$ the vector spaces that span over the basis elements
$e^{a_1\ldots a_k}$. Elements of $\cl_k(p,q)$ are said to be
elements of rank $k$. We have
\begin{eqnarray}
\cl(p,q)=\bigoplus_{k=0}^{n}\cl_k(p,q).\label{ranks}
\end{eqnarray}

Also we have classification of Clifford algebra elements based on the notion of parity (Clifford algebra as a superalgebra):
\begin{eqnarray}
\cl(p,q)=\cl_{\Even}(p,q)\oplus\cl_{\Odd}(p,q),\label{evenness}
\end{eqnarray}
where
$$\cl_{\Even}(p,q)=\cl_0(p,q)\oplus\cl_2(p,q)\oplus\cl_4(p,q)\oplus\ldots$$
$$\cl_{\Odd}(p,q)=\cl_1(p,q)\oplus\cl_3(p,q)\oplus\cl_5(p,q)\oplus\ldots$$

Denote by $[U,V]$ the commutator and  by $\{U,V\}$ the anticommutator of
Clifford algebra elements
$$
[U,V]=UV-VU,\quad \{U,V\}=UV+VU.
$$

Let us consider the Clifford algebra $\cl^\R(p,q)$ as the vector space and represent it in the form of the direct sum of four subspaces of {\it quaternion types} 0, 1, 2 and 3 (see \cite{QuatAaca}, \cite{Quat2Aaca}, \cite{DAN}):
\begin{equation}
\cl^\R(p,q)=\cl^\R_{\overline 0}(p,q)\oplus\cl^\R_{\overline 1}(p,q)\oplus
\cl^\R_{\overline 2}(p,q)\oplus\cl^\R_{\overline 3}(p,q),\label{kv}
\end{equation}
where
\begin{eqnarray*}
\cl^\R_{\overline
0}(p,q)&=&\cl^\R_0(p,q)\oplus\cl^\R_4(p,q)\oplus\cl^\R_8(p,q)\oplus\ldots,\\
\cl^\R_{\overline
1}(p,q)&=&\cl^\R_1(p,q)\oplus\cl^\R_5(p,q)\oplus\cl^\R_9(p,q)\oplus\ldots,\\
\cl^\R_{\overline
2}(p,q)&=&\cl^\R_2(p,q)\oplus\cl^\R_6(p,q)\oplus\cl^\R_{10}(p,q)\oplus\ldots,\\
\cl^\R_{\overline
3}(p,q)&=&\cl^\R_3(p,q)\oplus\cl^\R_7(p,q)\oplus\cl^\R_{11}(p,q)\oplus\ldots
\end{eqnarray*}
and in the right hand parts there are direct sums of subspaces with dimensions differ on 4.

We denote $\cl^\R_{\overline k}(p,q)$ by $\overline{\textbf{k}}$ and have the following properties (see \cite{QuatAaca}, \cite{Quat2Aaca}, \cite{DAN})

\begin{eqnarray}
&&[\overline{\textbf{k}},\overline{\textbf{k}}]\subseteq\overline{\textbf{2}},\qquad k=0, 1, 2, 3 \nonumber;\\
&&[\overline{\textbf{k}},\overline{\textbf{2}}]\subseteq\overline{\textbf{k}}, \qquad k=0, 1, 2, 3 \label{1}; \\
&&[\overline{\textbf{0}},\overline{\textbf{1}}]\subseteq\overline{\textbf{3}}, \quad  [\overline{\textbf{0}},\overline{\textbf{3}}]\subseteq\overline{\textbf{1}}, \quad [\overline{\textbf{1}},\overline{\textbf{3}}]\subseteq\overline{\textbf{0}} \nonumber,
\end{eqnarray}
\begin{eqnarray}
&&\{\overline{\textbf{k}},\overline{\textbf{k}}\}\subseteq\overline{\textbf{0}},\qquad k=0, 1, 2, 3 \nonumber;\\
&&\{\overline{\textbf{k}},\overline{\textbf{0}}\}\subseteq\overline{\textbf{k}}, \qquad k=0, 1, 2, 3; \label{2} \\
&&\{\overline{\textbf{1}},\overline{\textbf{2}}\}\subseteq\overline{\textbf{3}},  \quad \{\overline{\textbf{1}},\overline{\textbf{3}}\}\subseteq\overline{\textbf{2}}, \quad \{\overline{\textbf{2}},\overline{\textbf{3}}\}\subseteq\overline{\textbf{1}}\nonumber.
\end{eqnarray}

We represent complex Clifford algebra $\cl(p,q)$ in the form of the direct sum of eight subspaces:
\begin{equation}
\cl(p,q)=\overline{\textbf{0}}\oplus\overline{\textbf{1}}\oplus\overline{\textbf{2}}\oplus\overline{\textbf{3}} \oplus i\overline{\textbf{0}}\oplus i\overline{\textbf{1}}\oplus i\overline{\textbf{2}}\oplus i\overline{\textbf{3}}
\end{equation}

Consider the following linear operations in $\cl(p,q)$:
$$
U^{\curlywedge}=U|_{e^a\to-e^a},\quad
U^\sim=U|_{e^{a_1\ldots a_r}\to e^{a_r}\ldots e^{a_1}}.
$$

The operation $U\to U^{\curlywedge}$ is called {\it grade involution} and $U\to U^{\sim}$ is called {\it reversion}.

Also we have operation of {\it complex conjugation}
\begin{eqnarray}
\bar U=\bar u e+\bar u_a e^a+\sum_{a_1<a_2}\bar u_{a_1 a_2}e^{a_1
a_2}+\sum_{a_1<a_2<a_3}\bar u_{a_1 a_2 a_3}e^{a_1 a_2
a_3}+\ldots\nonumber
\end{eqnarray}

Superposition of reverse and complex conjugation is {\em pseudo-hermitian conjugation}
$$
U^\ddagger=\bar U^\sim.
$$

\section{Matrix representations of real Clifford algebras in some cases}

We have the following well-known isomorphisms

\begin{equation}
\cl^\R(p,q)\simeq\left\lbrace
\begin{array}{ll}
\Mat(2^{\frac{n}{2}},\R), & \parbox{.5\linewidth}{ if $p-q\equiv0; 2\!\!\mod 8$;}\\
\Mat(2^{\frac{n-1}{2}},\R)\oplus \Mat(2^{\frac{n-1}{2}},\R), & \parbox{.5\linewidth}{ if $p-q\equiv1\!\!\mod 8$;}\\
\Mat(2^{\frac{n-1}{2}},\C), & \parbox{.5\linewidth}{ if $p-q\equiv3; 7\!\!\mod 8$;}\\
\Mat(2^{\frac{n-2}{2}},\mathbb H), & \parbox{.5\linewidth}{ if $p-q\equiv4; 6\!\!\mod 8$;}\\
\Mat(2^{\frac{n-3}{2}},\mathbb H)\oplus \Mat(2^{\frac{n-3}{2}},\mathbb H), & \parbox{.5\linewidth}{ if $p-q\equiv5\!\!\mod 8$.}
\end{array}
\right.\nonumber
\end{equation}

Let use the following matrix representations of real Clifford algebra in the cases $p-q=0, 1, 2\mod 4$.

\begin{itemize}
  \item In the case $\cl^\R(0,0)$: $e\to 1$.

  \item In the case $\cl^\R(1,0)$: $e\to \diag(1,1)$, $e^1\to \diag(1,-1)$.
\end{itemize}

For basis element $e^{a_1 \ldots a_k}$ we use matrix which equals the product of matrices which correspond to generators $e^{a_1}, \ldots, e^{a_k}$. For identity element $e$ we always use identity matrix.

Let we have the matrix representation $\beta$ of $\cl^\R(p,q)$
\begin{eqnarray}
e^a \to \beta^a,\quad a=1, \ldots, n.\label{matpred}
\end{eqnarray}

\begin{enumerate}
  \item Let consider $\cl^\R(p+1,q+1)$. If $p-q\neq 1\!\!\mod 4$, then for $p$ generators with squares equals $+1$ and $q$ generators with squares equals $-1$ we have
$$\left( \begin{array}{ll}
 \beta^a & 0 \\
 0 & -\beta^a \end{array}\right).$$
And for the last two generators we have
$$
 \beta^+ =\left( \begin{array}{ll}
 0 & {\bf 1} \\
 {\bf 1} & 0 \end{array}\right),\quad
 \beta^- =\left( \begin{array}{ll}
 0 & -{\bf 1} \\
 {\bf 1} & 0 \end{array}\right).
 $$

  \item If $p-q\equiv1\!\!\mod 4$. Then matrices (\ref{matpred}) are block-diagonal and we have the following matrix representation of $\cl^\R(p+1,q+1)$. For $p+q$ generators we have the same and for the last two generators we have
 $$
 \beta^- =\left( \begin{array}{ll}
 \beta^* & 0 \\
 0 & -\beta^* \end{array}\right),\quad
 \beta^+ =\left( \begin{array}{ll}
 \beta^1\ldots\beta^n\beta^* & 0 \\
 0 & -\beta^1\ldots\beta^n\beta^* \end{array}\right),
 $$
where
$$
\beta^* =\left( \begin{array}{ll}
 0 & -{\bf 1} \\
 {\bf 1} & 0 \end{array}\right).
$$

  \item Matrix representation of $\cl^\R(q+1,p-1)$ is the following
$$(e^1)'\to \beta^1,\qquad (e^i)'\to \beta^i \beta^1,\quad i=2, \ldots, n.$$
  \item Matrix representation of $\cl^\R(p-4,q+4)$ is the following
$$(e^i)'\to \beta^i \beta^1 \beta^2 \beta^3 \beta^4,\quad i=1, 2, 3, 4,\qquad (e^j)'=e^j,\quad j=5, \ldots, n.$$
\end{enumerate}

Let's give some examples.

\begin{description}
  \item[$\cl^\R(1,1)$]
$$
 e^1 \to \left( \begin{array}{ll}
 0 & 1 \\
 1 & 0 \end{array}\right),\quad
 e^2 \to \left( \begin{array}{ll}
 0 & -1 \\
 1 & 0 \end{array}\right).
 $$
  \item[$\cl^\R(2,0)$ ]
 $$
  e^1 \to \left( \begin{array}{ll}
 0 & 1 \\
 1 & 0 \end{array}\right),\quad
 e^2 \to \left( \begin{array}{ll}
 -1 & 0 \\
 0 & 1 \end{array}\right).
 $$
 \item[$\cl^\R(2,1)$]
 $$
  e^1 \to \left( \begin{array}{llll}
 1 & 0 & 0 & 0\\
 0 & -1 & 0 & 0\\
 0 & 0 & -1 & 0\\
 0 & 0 & 0 & 1\end{array}\right),\quad
  e^2 \to \left( \begin{array}{llll}
 0 & 1 & 0 & 0\\
 1 & 0 & 0 & 0\\
 0 & 0 & 0 & -1\\
 0 & 0 & -1 & 0\end{array}\right),$$
  $$ e^3 \to \left( \begin{array}{llll}
 0 & 1 & 0 & 0\\
 -1 & 0 & 0 & 0\\
 0 & 0 & 0 & -1\\
 0 & 0 & 1 & 0\end{array}\right).
 $$
\end{description}

Note that we have the following relation between operation of Hermitian conjugate $\dagger$ and other operations in complex Clifford algebra $\cl(p,q)$ (see \cite{Marchuk:Shirokov})
\begin{eqnarray}
U^\dagger&=&e_{1\ldots p}U^\ddagger e^{1 \ldots p},\qquad \mbox{if $p$ - odd},\nonumber\\
U^\dagger&=&e_{1\ldots p}U^{\ddagger\curlywedge} e^{1 \ldots p},\qquad \mbox{if $p$ - even},\label{sogldager2}\\
U^\dagger&=&e_{p+1\ldots n}U^\ddagger e^{p+1 \ldots n},\qquad \mbox{if $q$ - even},\nonumber\\
U^\dagger&=&e_{p+1\ldots n}U^{\ddagger\curlywedge} e^{p+1 \ldots n},\qquad \mbox{if $q$ - odd}.\nonumber
\end{eqnarray}

For example, $(e^a)^\dagger=\eta^{aa}e^a=(e^a)^{-1}$. This operation corresponds to Hermitian conjugation of matrix:  $\beta^\dagger(U)=\beta(U^\dagger)$ (for matrix representation $\beta$).

So, in the cases $p-q=0, 1, 2\mod 8$ for the real Clifford algebra $\cl^\R(p,q)\subset\cl(p,q)$ we obtain
\begin{eqnarray}
U^T&=&e_{1 \ldots p} U^\sim e^{1 \ldots p},\qquad \mbox{if $p$ - odd},\nonumber\\
U^T&=&e_{1 \ldots p} U^{\sim\curlywedge} e^{1 \ldots p},\qquad \mbox{if $p$ - even},\label{sogltransp33}\\
U^T&=&e_{p+1 \ldots n} U^\sim e^{p+1 \ldots n},\qquad \mbox{if $q$ - even},\nonumber\\
U^T&=&e_{p+1 \ldots n} U^{\sim\curlywedge} e^{p+1 \ldots n},\qquad \mbox{if $q$ - odd,}\nonumber
\end{eqnarray}
where $U^T$ is transpose of $U$ (of corresponding matrix representation).

\begin{thm} Consider real Clifford algebra $\cl^\R(p,q)$ of signatures
\begin{eqnarray}
p-q\equiv 0, 1, 2\mod 8.\nonumber
\end{eqnarray}
Then there exists such matrix representation that $\gamma: e^a \to \gamma^a$ - real matrices and
$$(\gamma^a)^\dagger=(\gamma^a)^T=\eta^{aa}\gamma^a,\quad a=1,\ldots, n,$$
i.e. $\gamma(U^\dagger)=\gamma^\dagger(U)=\gamma^T(U) \quad\forall U\in\cl^\R(p,q)$.

Moreover,
\begin{itemize}
  \item in the case of even $n$, $p\neq 0$
$$
\gamma^{1\ldots p}=\left\lbrace
\begin{array}{ll}
\Omega, & \mbox{\rm if $p=2, 3\mod 4$,}\\
J, & \mbox{\rm if $p=0, 1\mod 4$.}
\end{array}
\right.
$$

  \item in the case of even $n$, $q\neq 0$
$$
\gamma^{p+1\ldots n}=\left\lbrace
\begin{array}{ll}
\Omega, & \mbox{\rm if $q=1, 2\mod 4$,}\\
J, & \mbox{\rm if $q=0, 3 \mod 4$.}
\end{array}
\right.
$$

\item in the case of odd $n\geq 3$,  $p\neq 0$ - even ($q$ - odd)
$$
 \gamma^{1\ldots p}=\left\lbrace
\begin{array}{ll}
\diag(\Omega, \Omega), & \mbox{\rm if $p=2\mod 4$,}\\
\diag(J, J), & \mbox{\rm if $p=0\mod 4$.}
\end{array}
\right.
$$

Moreover, each of block-diagonal matrices $\gamma^a$ consists of two identical blocks with different signs.

  \item in the case of odd $n\geq 3$, $q\neq 0$ - even ($p$ - odd)
$$
 \gamma^{p+1\ldots n}=\left\lbrace
\begin{array}{ll}
\diag(\Omega, \Omega), & \mbox{\rm if $q=2\mod 4$,}\\
\diag(J, J), & \mbox{\rm if $q=0\mod 4$.}
\end{array}
\right.
$$

Moreover, each of block-diagonal matrices $\gamma^a$ consists of two identical blocks with different signs.
  \end{itemize}

Here  $\Omega$ is the block matrix
$$\Omega=\left( \begin{array}{ll}
 0 & -{\bf1} \\
 {\bf1} & 0 \end{array}\right)$$
 and $J=\diag(1, \ldots, 1, -1, \ldots ,-1)$ is the diagonal matrix with the same number of $1$ and $-1$ on the diagonal.
\end{thm}

\proof.\, We use matrix representation which has been discussed before the theorem. For this representation $\beta$ we have $(\beta^a)^\dagger=\eta^{aa}\beta^a,\quad a=1,\ldots, n.$

Note that $\Omega^T=\Omega$, $\Omega^2=-1$, $\tr\Omega=0$ and spectrum of matrix $\Omega$ consists of the same number of $i$ and $-i$.
For the matrix $J$ we have $J^T=J$, $J^2=1$, $\tr J=0$ and spectrum of matrix $J$ consists of the same number of $1$ and $-1$.

Now let consider real matrix $N=\beta^{1\ldots p}$. We have $N^T=N^\dagger=N^{-1}$, $\tr N=0$.

In the cases $p\equiv 2, 3\mod 4$ we obtain $N^2={\bf - 1}$. So, the spectrum of matrix $N$ coincides with the spectrum of matrix $\Omega$. Matrices $\Omega$ and $N$ are orthogonal, they are reduced to the same diagonal form. So there exists orthogonal matrix $T^T=T^{-1}$ such that $\Omega=T^{-1}NT$. Now we consider transformation $T^{-1}\beta^a T=\gamma^a$ and obtain another matrix representation of Clifford algebra, moreover $\gamma^{1\ldots p}=\Omega$. Matrix $T$ is real, then matrices $\gamma^a$ are real too. Since $T^\dagger=T^{-1}$, then we have $(\gamma^a)^\dagger=\eta^{aa}\gamma^a$.

In the cases $p\equiv 0, 1\mod 4$ we have $N^2={\bf 1}$. Thus the spectrum of matrix $N$ coincides with the spestrum of matrix $J$. The proof is analogous.

The second statement of the theorem has the analogous proof.

Now let consider the case of odd $n$. We use matrix representation $\beta$ again. Since $p$ is even, then the matrix $\beta^{1\ldots p}=\diag(D,D)$ consists of the two identical blocks $D$. Further we reduce each of the blocks to the $\Omega$ (or $J$). We obtain matrices $\gamma^{a}$ that consist of two identical blocks with different signs.

The last statement has the analogous proof. $\blacksquare$

\section{The main theorems}

Let consider the following subsets of Clifford algebra
\begin{eqnarray}
\Sp\O^\R_{2i1}\cl(p,q) &=& \{U\in\cl^\R_\Even(p,q)\oplus i\cl^\R_\Odd(p,q) : U^\ddagger U=e\},\nonumber\\
\Sp\O^\R_{2i3}\cl(p,q) &=& \{U\in\cl^\R_\Even(p,q)\oplus i\cl^\R_\Odd(p,q) : U^{\ddagger\curlywedge} U=e\},\nonumber\\
\Sp\O^\R_{23}\cl(p,q)&=&\{U\in\cl^\R(p,q)\, | \, U^{\sim}U=e\},\label{grou}\\
\Sp\O^\R_{12}\cl(p,q)&=&\{U\in\cl^\R(p,q)\, | \, U^{\sim\curlywedge}U=e\},\nonumber\\
\Sp\O^\R_{2}\cl(p,q)&=&\{U\in\cl^\R_{\Even}(p,q)\, | \, U^{\sim}U=e\}.\nonumber
\end{eqnarray}
They can be considered as Lie groups. Their Lie algebras are
\begin{eqnarray}
\overline{\textbf{2}}\oplus i\overline{\textbf{1}},\quad \overline{\textbf{2}}\oplus i\overline{\textbf{3}},\quad \overline{\textbf{23}},\quad \overline{\textbf{12}},\quad \overline{\textbf{2}}\label{algebr}
\end{eqnarray}
respectively.

Note that spin group
$$\Spin_{+}(p,q)=\{U\in\cl^\R_{\Even}(p,q) | \forall x\in\cl^\R_1(p,q), UxU^{-1}\in\cl^\R_1(p,q), U^{\sim}U=e\}$$
is a subgroup of all 5 considered Lie groups (\ref{grou}). Moreover, spin group $\Spin_{+}(p,q)$ coincides with group $\Sp\O^\R_{2}\cl(p,q)$ in the case of dimensions $n\leq 5$. Lie algebra $\cl_2^\R(p,q)$ of Lie group $\Spin_{+}(p,q)$ is a subalgebra of algebras (\ref{algebr}). Moreover, Lie algebra $\cl_2^\R(p,q)$ coincides with Lie algebra $\overline{\textbf{2}}$ in the cases of dimensions $n\leq 5$, because notions of rank and quaternion type coincide in these cases.

\begin{thm}\label{theoremggg} We have the following Lie group isomorphisms
$$\Sp\O^\R_{2i1}\cl(p,q)\simeq \Sp\O^\R_{12}\cl(q,p),$$
$$\Sp\O^\R_{2i3}\cl(p,q)\simeq \Sp\O^\R_{23}\cl(q,p),$$
$$\Sp\O^\R_{2}\cl(p,q)\simeq \Sp\O^\R_{12}\cl(p,q-1)\simeq \Sp\O^\R_{12}\cl(q,p-1),$$
$$\Sp\O^\R_{2}\cl(p,q)\simeq \Sp\O^\R_{2}\cl(q,p).$$
\end{thm}

\proof.\, We must consider transformation $e^a \to e^a e^n$ or $e^a \to ie^a$ in different cases. $\blacksquare$

\begin{thm} We have the following Lie group isomorphisms.

In the cases of signatures
$p-q\equiv 0, 1, 2\mod 8$
$$\Sp\O^\R_{23}\cl(p,q)=\{U\in\cl^\R(p,q)\, | \, U^{\sim}U=e\}\simeq$$
\begin{equation}
\left\lbrace
\begin{array}{ll}
\O(2^{\frac{n}{2}}), & \mbox{\rm $(p,q)=(n,0)$, $n$ is even;}\\
\O(2^{\frac{n-1}{2}})\times\O(2^{\frac{n-1}{2}}), & \mbox{\rm $(p,q)=(n,0)$, $n$ is odd;}\\
\O(2^{\frac{n}{2}-1},2^{\frac{n}{2}-1}), & \parbox{.5\linewidth}{$n\equiv 0, 2\!\!\mod 8$, $q\neq 0$;}\\
\Sp(2^{\frac{n}{2}-1},\R), & \parbox{.5\linewidth}{$n\equiv4, 6\!\!\mod 8$;}\\
\O(2^{\frac{n-1}{2}-1},2^{\frac{n-1}{2}-1})\times \O(2^{\frac{n-1}{2}-1},2^{\frac{n-1}{2}-1}), & \parbox{.5\linewidth}{$n\equiv 1\!\!\mod 8$, $q\neq 0$;}\\
\Sp(2^{\frac{n-1}{2}-1},\R)\times \Sp(2^{\frac{n-1}{2}-1},\R), & \parbox{.5\linewidth}{$n\equiv 5\!\!\mod 8$;}\\
\GL(2^{\frac{n-1}{2}},\R), & \parbox{.5\linewidth}{$n\equiv 3, 7\!\!\mod 8$.}
\end{array}
\right.\nonumber
\end{equation}
In the cases of signatures
$p-q\equiv 0, 1, 2\mod 8$
$$\Sp\O^\R_{12}\cl(p,q)=\{U\in\cl^\R(p,q)\, | \, U^{\sim\curlywedge}U=e\}\simeq$$
\begin{equation}
\left\lbrace
\begin{array}{ll}
\O(2^{\frac{n}{2}}), & \mbox{\rm $(p,q)=(0,n)$, $n$ is even;}\\
\O(2^{\frac{n-1}{2}})\times\O(2^{\frac{n-1}{2}}), & \mbox{\rm $(p,q)=(0,n)$, $n$ is odd;}\\
\O(2^{\frac{n}{2}-1},2^{\frac{n}{2}-1}), & \parbox{.5\linewidth}{$n\equiv 0, 6\!\!\mod 8$, $p\neq 0$;}\\
\Sp(2^{\frac{n}{2}-1},\R), & \parbox{.5\linewidth}{$n\equiv2, 4\!\!\mod 8$;}\\
\O(2^{\frac{n-1}{2}-1},2^{\frac{n-1}{2}-1})\times \O(2^{\frac{n-1}{2}-1},2^{\frac{n-1}{2}-1}), & \parbox{.5\linewidth}{$n\equiv 7\!\!\mod 8$, $p\neq 0$;}\\
\Sp(2^{\frac{n-1}{2}-1},\R)\times \Sp(2^{\frac{n-1}{2}-1},\R), & \parbox{.5\linewidth}{$n\equiv 3\!\!\mod 8$;}\\
\GL(2^{\frac{n-1}{2}},\R), & \parbox{.5\linewidth}{$n\equiv 1, 5\!\!\mod 8$.}
\end{array}
\right.\nonumber
\end{equation}
In the cases of signatures
$p-q=0, 6, 7\mod 8$
$$\Sp\O^\R_{2i1}\cl(p,q)=\{U\in\cl^\R_\Even(p,q)\oplus i\cl^\R_\Odd(p,q) : U^\ddagger U=e\}\simeq$$
\begin{equation}
\left\lbrace
\begin{array}{ll}
\O(2^{\frac{n}{2}}), & \mbox{\rm $(p,q)=(n,0)$, $n$ is even;}\\
\O(2^{\frac{n-1}{2}})\times\O(2^{\frac{n-1}{2}}), & \mbox{\rm $(p,q)=(n,0)$, $n$ is odd;}\\
\O(2^{\frac{n}{2}-1},2^{\frac{n}{2}-1}), & \parbox{.5\linewidth}{$n\equiv 0, 6\!\!\mod 8$, $q\neq 0$;}\\
\Sp(2^{\frac{n}{2}-1},\R), & \parbox{.5\linewidth}{$n\equiv2, 4\!\!\mod 8$;}\\
\O(2^{\frac{n-1}{2}-1},2^{\frac{n-1}{2}-1})\times \O(2^{\frac{n-1}{2}-1},2^{\frac{n-1}{2}-1}), & \parbox{.5\linewidth}{$n\equiv 7\!\!\mod 8$, $q\neq 0$;}\\
\Sp(2^{\frac{n-1}{2}-1},\R)\times \Sp(2^{\frac{n-1}{2}-1},\R), & \parbox{.5\linewidth}{$n\equiv 3\!\!\mod 8$;}\\
\GL(2^{\frac{n-1}{2}},\R), & \parbox{.5\linewidth}{$n\equiv 1, 5\!\!\mod 8$.}
\end{array}
\right.\nonumber
\end{equation}
In the cases of signatures
$p-q\equiv 0, 6, 7 \mod 8$
$$\Sp\O^\R_{2i3}\cl(p,q)=\{U\in\cl^\R_\Even(p,q)\oplus i\cl^\R_\Odd(p,q) : U^{\ddagger\curlywedge} U=e\}\simeq$$
\begin{equation}
\left\lbrace
\begin{array}{ll}
\O(2^{\frac{n}{2}}), & \mbox{\rm $(p,q)=(0,n)$, $n$ is even;}\\
\O(2^{\frac{n-1}{2}})\times\O(2^{\frac{n-1}{2}}), & \mbox{\rm $(p,q)=(0,n)$, $n$ is odd;}\\
\O(2^{\frac{n}{2}-1},2^{\frac{n}{2}-1}), & \parbox{.5\linewidth}{$n\equiv 0, 2\!\!\mod 8$, $p\neq 0$;}\\
\Sp(2^{\frac{n}{2}-1},\R), & \parbox{.5\linewidth}{$n\equiv4, 6\!\!\mod 8$;}\\
\O(2^{\frac{n-1}{2}-1},2^{\frac{n-1}{2}-1})\times \O(2^{\frac{n-1}{2}-1},2^{\frac{n-1}{2}-1}), & \parbox{.5\linewidth}{$n\equiv 1\!\!\mod 8$, $p\neq 0$;}\\
\Sp(2^{\frac{n-1}{2}-1},\R)\times \Sp(2^{\frac{n-1}{2}-1},\R), & \parbox{.5\linewidth}{$n\equiv 5\!\!\mod 8$;}\\
\GL(2^{\frac{n-1}{2}},\R), & \parbox{.5\linewidth}{$n\equiv 3, 7\!\!\mod 8$.}
\end{array}
\right.\nonumber
\end{equation}
In the cases of signatures
$p-q = 0, 1, 7\mod 8$
$$\Sp\O^\R_{2}\cl(p,q)=\{U\in\cl^\R_{\Even}(p,q)\, | \, U^{\sim}U=e\}\simeq$$
\begin{equation}
\left\lbrace
\begin{array}{ll}
\O(2^{\frac{n-1}{2}}), & \mbox{\rm $(p,q)=(n,0), (0,n)$, $n$ is odd;}\\
\O(2^{\frac{n-2}{2}})\times \O(2^{\frac{n-2}{2}}), & \mbox{\rm $(p,q)=(n,0), (0,n)$, $n$ is even;}\\
\O(2^{\frac{n-1}{2}-1},2^{\frac{n-1}{2}-1}), & \parbox{.5\linewidth}{$n\equiv 1, 7\!\!\mod 8$;}\\
\Sp(2^{\frac{n-1}{2}-1},\R), & \parbox{.5\linewidth}{$n\equiv3, 5\!\!\mod 8$;}\\
\O(2^{\frac{n-2}{2}-1},2^{\frac{n-2}{2}-1})\times \O(2^{\frac{n-2}{2}-1},2^{\frac{n-2}{2}-1}), & \parbox{.5\linewidth}{$n\equiv 0\!\!\mod 8$;}\\
\Sp(2^{\frac{n-2}{2}-1},\R)\times \Sp(2^{\frac{n-2}{2}-1},\R), & \parbox{.5\linewidth}{$n\equiv 4\!\!\mod 8$;}\\
\GL(2^{\frac{n-2}{2}},\R), & \parbox{.5\linewidth}{$n\equiv 2, 6\!\!\mod 8$.}
\end{array}
\right.\nonumber
\end{equation}
\end{thm}

We have
$$\GL(n, \R)=\{A\in\Mat(n,\R) \,|\, \exists A^{-1}\},$$
$$\O(n)=\{A\in \Mat(n,\R)\,|\, A^T A={\bf1}\},\,\O(p,q)=\{A\in \Mat(n,\R)\,|\, A^T \eta A=\eta\},$$
$$\Sp(n, \R)=\{A\in\Mat(2n, \R): A^T\Omega A=\Omega\}.$$

\proof. \, At first let's prove the following isomorphisms for the group $\Sp\O^\R_{23}\cl(p,q)$:
$$\Sp\O^\R_{23}\cl(p,q)\simeq$$
\begin{eqnarray}
\left\lbrace
\begin{array}{ll}
\O(2^{\frac{n}{2}}), & \mbox{\rm $(p,q)=(n,0)$, $n$ is even;}\\
\O(2^{\frac{n-1}{2}})\times\O(2^{\frac{n-1}{2}}), & \mbox{\rm $(p,q)=(n,0)$, $n$ is odd;}\\
\O(2^{\frac{n}{2}-1},2^{\frac{n}{2}-1}), & \parbox{.5\linewidth}{\rm $(p, q)=(\underline{0}, \underline{0}), (\underline{2}, \underline{0}), (\underline{1}, \underline{1}), (\underline{1}, \underline{3})$,\\ where $q\neq 0$;}\\
\O(2^{\frac{n-1}{2}-1},2^{\frac{n-1}{2}-1})\times \O(2^{\frac{n-1}{2}-1},2^{\frac{n-1}{2}-1}), & \mbox{\rm $(p, q)=(\underline{1}, \underline{0})$, where $q\neq 0$;}\\
\Sp(2^{\frac{n}{2}-1},\R), & \mbox{\rm $(p, q)=(\underline{3}, \underline{1}), (\underline{3}, \underline{3}), (\underline{0}, \underline{2}), (\underline{2}, \underline{2})$;}\\
\Sp(2^{\frac{n-1}{2}-1},\R)\times \Sp(2^{\frac{n-1}{2}-1},\R), & \mbox{\rm $(p, q)=(\underline{3}, \underline{2})$;}\\
\GL(2^{\frac{n-1}{2}},\R), & \mbox{\rm $(p, q)=(\underline{2}, \underline{1}), (\underline{0}, \underline{3})$.}
\end{array}
\right.\nonumber
\end{eqnarray}
where $\underline{k}$ means $k$ modulo 4.

We use real matrix representation ($\gamma^a$ are real).

In the case of signature $(n,0)$ we have $U^\sim=U^\ddagger=U^\dagger=U^T$ and obtain isomorphisms from the statement.

Consider case $q\neq 0$. We use matrix representation from Theorem 3.1. For the odd $p$ and $q$ we have $U^\dagger=e_{1\ldots p} U^\ddagger e^{1\ldots p}$, for the even $p$ and $q$ we heve $U^\dagger=e_{p+1\ldots n} U^\ddagger e^{p+1\ldots n}$.  In both cases we also have $U^\dagger=U^T$ and $U^\ddagger=U^\sim$. Then we obtain isomorphisms between considered groups and $\Sp(2^{\frac{n}{2}-1},\R)$ or $\O(2^{\frac{n}{2}-1},2^{\frac{n}{2}-1})$ in different cases.

In the case of odd $p$ and even $q$ we have $U^\dagger=e_{p+1\ldots n} U^\ddagger e^{p+1\ldots n}$, matrix representation is block-diagonal.

In the case of even $p\neq 0$ and odd $q$ we have $U^\dagger=e_{1\ldots p} U^{\ddagger\curlywedge} e^{1\ldots p}$.
The even part of arbitrary Clifford algebra element has the form $\diag(A,A)$ and the odd part of element has the form $\diag(B,-B)$. Then we obtain
$$(\diag(A-B,A+B))^T \diag(G, G) \diag(A+B,A-B)=\diag(G, G).$$ So, we have $$ (A-B)^T G(A+B)=G,$$
where $G$ is $\Omega$ or $J$ in different cases. In these cases we obtain the isomorphism with the group $\GL(2^{\frac{n-1}{2}},\R)$.

In the case $p=0$, $q$ - odd we have $U^\dagger=U^{\ddagger\curlywedge}$ and obtain $U^{\dagger\curlywedge}U=e$. Then
$$(\diag(A-B,A+B))^T \diag(A+B,A-B) = {\bf 1}$$
and also we obtain isomorphism with linear group.

It's easy to rewrite these statements in the form as in the statement of the theorem.

We can obtain isomorphisms for the group $\Sp\O^\R_{12}\cl(p,q)$ analogously: in the cases
$p-q\equiv 0, 1, 2\mod 8$ we have
$$\Sp\O^\R_{12}\cl(p,q)\simeq$$
\begin{eqnarray}
\left\lbrace
\begin{array}{ll}
\O(2^{\frac{n}{2}}), & \mbox{\rm $(p,q)=(0,n)$, $n$ is even;}\\
\O(2^{\frac{n-1}{2}})\times\O(2^{\frac{n-1}{2}}), & \mbox{\rm $(p,q)=(0,n)$, $n$ is odd;}\\
\O(2^{\frac{n}{2}-1},2^{\frac{n}{2}-1}), & \parbox{.5\linewidth}{\rm $(p, q)=(\underline{0}, \underline{0}), (\underline{0}, \underline{2}), (\underline{1}, \underline{3}), (\underline{3}, \underline{3})$,\\ where $p\neq 0$;}\\
\O(2^{\frac{n-1}{2}-1},2^{\frac{n-1}{2}-1})\times \O(2^{\frac{n-1}{2}-1},2^{\frac{n-1}{2}-1}), & \mbox{\rm $(p, q)=(\underline{0}, \underline{3})$, where $p\neq 0$;}\\
\Sp(2^{\frac{n}{2}-1},\R), & \mbox{\rm $(p, q)=(\underline{1}, \underline{1}), (\underline{3}, \underline{1}), (\underline{2}, \underline{0}), (\underline{2}, \underline{2})$;}\\
\Sp(2^{\frac{n-1}{2}-1},\R)\times \Sp(2^{\frac{n-1}{2}-1},\R), & \mbox{\rm $(p, q)=(\underline{2}, \underline{1})$;}\\
\GL(2^{\frac{n-1}{2}},\R), & \mbox{\rm $(p, q)=(\underline{1}, \underline{0}), (\underline{3}, \underline{2})$.}
\end{array}
\right.\nonumber
\end{eqnarray}
where $\underline{k}$ means $k$ modulo 4.

We can obtain statements for the groups $\Sp\O^\R_{23}\cl(p,q)$, $\Sp\O^\R_{12}\cl(p,q)$, $\Sp\O^\R_{2}\cl(p,q)$ with the use of Theorem 4.1. $\blacksquare$

We have the following definitions of classical Lie algebras (as matrix algebras with operation of commutator):
$$\gl(n, \R)=\Mat(n,\R),\quad \o(n)=\{A\in \Mat(n,\R)\,|\, A^T=-A\},$$
$$\o(p,q)=\{A\in \Mat(n,\R)\,|\, A^T \eta =-\eta A\},$$
$$\sp(n,\R)=\{A\in \Mat(2n,\R)\,|\, A^T \Omega =- \Omega A\}.$$

We obtain the following Lie algebra isomorphisms.

In the cases of signatures
$p-q\equiv 0, 1, 2\mod 8$
\begin{equation}
\overline{\textbf{23}}\simeq\left\lbrace
\begin{array}{ll}
\o(2^{\frac{n}{2}}), & \mbox{\rm $(p,q)=(n,0)$, $n$ is even;}\\
\o(2^{\frac{n-1}{2}})\oplus\o(2^{\frac{n-1}{2}}), & \mbox{\rm $(p,q)=(n,0)$, $n$ is odd;}\\
\o(2^{\frac{n}{2}-1},2^{\frac{n}{2}-1}), & \parbox{.5\linewidth}{$n\equiv 0, 2\!\!\mod 8$, $q\neq 0$;}\\
\sp(2^{\frac{n}{2}-1},\R), & \parbox{.5\linewidth}{$n\equiv4, 6\!\!\mod 8$;}\\
\o(2^{\frac{n-1}{2}-1},2^{\frac{n-1}{2}-1})\oplus \o(2^{\frac{n-1}{2}-1},2^{\frac{n-1}{2}-1}), & \parbox{.5\linewidth}{$n\equiv 1\!\!\mod 8$, $q\neq 0$;}\\
\sp(2^{\frac{n-1}{2}-1},\R)\oplus \sp(2^{\frac{n-1}{2}-1},\R), & \parbox{.5\linewidth}{$n\equiv 5\!\!\mod 8$;}\\
\gl(2^{\frac{n-1}{2}},\R), & \parbox{.5\linewidth}{$n\equiv 3, 7\!\!\mod 8$.}
\end{array}
\right.\nonumber
\end{equation}
In the cases of signatures
$p-q\equiv 0, 1, 2\mod 8$
\begin{equation}
\overline{\textbf{12}}\simeq\left\lbrace
\begin{array}{ll}
\o(2^{\frac{n}{2}}), & \mbox{\rm $(p,q)=(0,n)$, $n$ is even;}\\
\o(2^{\frac{n-1}{2}})\oplus\o(2^{\frac{n-1}{2}}), & \mbox{\rm $(p,q)=(0,n)$, $n$ is odd;}\\
\o(2^{\frac{n}{2}-1},2^{\frac{n}{2}-1}), & \parbox{.5\linewidth}{$n\equiv 0, 6\!\!\mod 8$, $p\neq 0$;}\\
\sp(2^{\frac{n}{2}-1},\R), & \parbox{.5\linewidth}{$n\equiv2, 4\!\!\mod 8$;}\\
\o(2^{\frac{n-1}{2}-1},2^{\frac{n-1}{2}-1})\oplus \o(2^{\frac{n-1}{2}-1},2^{\frac{n-1}{2}-1}), & \parbox{.5\linewidth}{$n\equiv 7\!\!\mod 8$, $p\neq 0$;}\\
\sp(2^{\frac{n-1}{2}-1},\R)\oplus \sp(2^{\frac{n-1}{2}-1},\R), & \parbox{.5\linewidth}{$n\equiv 3\!\!\mod 8$;}\\
\gl(2^{\frac{n-1}{2}},\R), & \parbox{.5\linewidth}{$n\equiv 1, 5\!\!\mod 8$.}
\end{array}
\right.\nonumber
\end{equation}
In the cases of signatures
$p-q=0, 6, 7\mod 8$
\begin{equation}
\overline{\textbf{2}}\oplus i\overline{\textbf{1}}\simeq\left\lbrace
\begin{array}{ll}
\o(2^{\frac{n}{2}}), & \mbox{\rm $(p,q)=(n,0)$, $n$ is even;}\\
\o(2^{\frac{n-1}{2}})\oplus\o(2^{\frac{n-1}{2}}), & \mbox{\rm $(p,q)=(n,0)$, $n$ is odd;}\\
\o(2^{\frac{n}{2}-1},2^{\frac{n}{2}-1}), & \parbox{.5\linewidth}{$n\equiv 0, 6\!\!\mod 8$, $q\neq 0$;}\\
\sp(2^{\frac{n}{2}-1},\R), & \parbox{.5\linewidth}{$n\equiv2, 4\!\!\mod 8$;}\\
\o(2^{\frac{n-1}{2}-1},2^{\frac{n-1}{2}-1})\oplus \o(2^{\frac{n-1}{2}-1},2^{\frac{n-1}{2}-1}), & \parbox{.5\linewidth}{$n\equiv 7\!\!\mod 8$, $q\neq 0$;}\\
\sp(2^{\frac{n-1}{2}-1},\R)\oplus \sp(2^{\frac{n-1}{2}-1},\R), & \parbox{.5\linewidth}{$n\equiv 3\!\!\mod 8$;}\\
\gl(2^{\frac{n-1}{2}},\R), & \parbox{.5\linewidth}{$n\equiv 1, 5\!\!\mod 8$.}
\end{array}
\right.\nonumber
\end{equation}
In the cases of signatures
$p-q\equiv 0, 6, 7 \mod 8$
\begin{equation}
\overline{\textbf{2}}\oplus i\overline{\textbf{3}}\simeq\left\lbrace
\begin{array}{ll}
\o(2^{\frac{n}{2}}), & \mbox{\rm $(p,q)=(0,n)$, $n$ is even;}\\
\o(2^{\frac{n-1}{2}})\oplus\o(2^{\frac{n-1}{2}}), & \mbox{\rm $(p,q)=(0,n)$, $n$ is odd;}\\
\o(2^{\frac{n}{2}-1},2^{\frac{n}{2}-1}), & \parbox{.5\linewidth}{$n\equiv 0, 2\!\!\mod 8$, $p\neq 0$;}\\
\sp(2^{\frac{n}{2}-1},\R), & \parbox{.5\linewidth}{$n\equiv4, 6\!\!\mod 8$;}\\
\o(2^{\frac{n-1}{2}-1},2^{\frac{n-1}{2}-1})\oplus \o(2^{\frac{n-1}{2}-1},2^{\frac{n-1}{2}-1}), & \parbox{.5\linewidth}{$n\equiv 1\!\!\mod 8$, $p\neq 0$;}\\
\sp(2^{\frac{n-1}{2}-1},\R)\oplus \sp(2^{\frac{n-1}{2}-1},\R), & \parbox{.5\linewidth}{$n\equiv 5\!\!\mod 8$;}\\
\gl(2^{\frac{n-1}{2}},\R), & \parbox{.5\linewidth}{$n\equiv 3, 7\!\!\mod 8$.}
\end{array}
\right.\nonumber
\end{equation}
In the cases of signatures
$p-q = 0, 1, 7\mod 8$
\begin{equation}
\overline{\textbf{2}}\simeq\left\lbrace
\begin{array}{ll}
\o(2^{\frac{n-1}{2}}), & \mbox{\rm $(p,q)=(n,0), (0,n)$, $n$ is odd;}\\
\o(2^{\frac{n-2}{2}})\oplus \o(2^{\frac{n-2}{2}}), & \mbox{\rm $(p,q)=(n,0), (0,n)$, $n$ is even;}\\
\o(2^{\frac{n-1}{2}-1},2^{\frac{n-1}{2}-1}), & \parbox{.5\linewidth}{$n\equiv 1, 7\!\!\mod 8$;}\\
\sp(2^{\frac{n-1}{2}-1},\R), & \parbox{.5\linewidth}{$n\equiv3, 5\!\!\mod 8$;}\\
\o(2^{\frac{n-2}{2}-1},2^{\frac{n-2}{2}-1})\oplus \o(2^{\frac{n-2}{2}-1},2^{\frac{n-2}{2}-1}), & \parbox{.5\linewidth}{$n\equiv 0\!\!\mod 8$;}\\
\sp(2^{\frac{n-2}{2}-1},\R)\oplus \sp(2^{\frac{n-2}{2}-1},\R), & \parbox{.5\linewidth}{$n\equiv 4\!\!\mod 8$;}\\
\gl(2^{\frac{n-2}{2}},\R), & \parbox{.5\linewidth}{$n\equiv 2, 6\!\!\mod 8$.}
\end{array}
\right.\nonumber
\end{equation}

\bigskip

In this paper we present some isomorphisms between Lie groups in Clifford algebra formalism and classical matrix groups. We obtain statements for corresponding Lie algebras. We discuss the relation between these Lie groups and spin group.

Note that in the cases of other signatures (not considered in Theorem 4.2) groups
$$\Sp\O^\R_{2i1}\cl(p,q),\,
\Sp\O^\R_{2i3}\cl(p,q),\,
\Sp\O^\R_{23}\cl(p,q),\,
\Sp\O^\R_{12}\cl(p,q),\,
\Sp\O^\R_{2}\cl(p,q)$$
are isomorphic not to the real matrix classical groups but to some complex and quaternion matrix groups. This is a subject for further research.

\subsection*{Acknowledgment}

This work was supported by Russian Science Foundation (project 14-11-00687, Steklov Mathematical Institute).

\end{document}